\documentclass[showpacs,preprintnumbers,
amsmath,amssymb,aps,prd]{revtex4}
\usepackage{}
\usepackage{amsfonts}
\usepackage{amsfonts,graphics,epsfig,subfigure}

\begin{document}

\title{Revisiting van der Waals like behavior of f(R) AdS black holes via the two point correlation function}

\author{Jie-Xiong Mo $^{a,b}$ \footnote{mojiexiong@gmail.com}, Gu-Qiang Li $^{a,b}$ \footnote{zsgqli@hotmail.com}, Ze-Tao Lin $^{b}$ \footnote{Albertlyn@163.com},Xiao-Xiong Zeng $^{c,d}$\footnote{xxzeng@itp.ac.cn}}

 \affiliation{$^a$ Institute of Theoretical Physics, Lingnan Normal University, Zhanjiang, 524048, Guangdong, China\\
 $^b$ Department of Physics, Lingnan Normal University, Zhanjiang, 524048, Guangdong, China\\
$^c$ School of Material Science and Engineering, Chongqing Jiaotong University, Chongqing, 400074, China\\
$^d$ Institute of Theoretical Physics, Chinese Academy of Sciences, Beijing 100190, China}

\begin{abstract}

Van der Waals like behavior of $f(R)$ AdS black holes is revisited via two point correlation function, which is dual to the geodesic length in the bulk. The equation of motion constrained by the boundary condition is solved numerically and both the effect of boundary region size and $f(R)$ gravity are probed. Moreover, an analogous specific heat related to $\delta L$ is introduced. It is shown that the $T-\delta L$ graphs of $f(R)$ AdS black holes exhibit reverse van der Waals like behavior just as the $T-S$ graphs do. Free energy analysis is carried out to determine the first order phase transition temperature $T_*$ and the unstable branch in $T-\delta L$ curve is removed by a bar $T=T_*$. It is shown that the first order phase transition temperature is the same at least to the order of $10^{-10}$ for different choices of the parameter $b$ although the values of free energy vary with $b$. Our result further supports the former finding that charged $f(R)$ AdS black holes behave much like RN-AdS black holes. We also check the analogous equal area law numerically and find that the relative errors for both the cases $\theta_0=0.1$ and $\theta_0=0.2$ are small enough. The fitting functions between $ \log\mid T -T_c\mid$ and $\log\mid\delta L-\delta L_c\mid $ for both cases are also obtained. It is shown that the slope is around 3, implying that the critical exponent is about $2/3$. This result is in accordance with those in former literatures of specific heat related to the thermal entropy or entanglement entropy.

\end{abstract}

\maketitle

\section{Introduction}
\label{sec:1}
  Van der Waals like behavior of black holes has long been an interesting topic in the black hole physics research for it discloses the close relation between black hole thermodynamics and ordinary thermodynamic systems. In the famous paper, Chamblin et al. \cite{Chamblin1,Chamblin2} found that Reissner-Nordstr\"{o}m-AdS (RN-AdS) black holes undergoes first order phase transition, which is analogous to the van der Waals liquid-gas
phase transition. Carlip and Vaidya investigated the thermodynamics of a four-dimensional charged black hole in
a finite cavity in asymptotically flat and asymptotically de Sitter space and discovered similar phase transition \cite{Carlip}. Lu et al. \cite{lujianxin} studied the phase structure of asymptotically flat nondilatonic as well as dilatonic black branes in a cavity in arbitrary dimensions. It was shown that the phase diagram has a line of first-order phase transition in a certain range of temperatures which ends up at a second order phase transition point when the charge is below a critical value \cite{lujianxin}. In this sense, van der Waals like behavior is such a universal phenomenon that it exists not only in AdS black holes, but also in asymptotically flat and asymptotically de Sitter black holes and black branes. Treating the cosmological constant as thermodynamic pressure, Kubiz\v{n}\'{a}k and Mann \cite{Kubiznak} investigated the $P$-$V$ criticality of RN-AdS black holes in the extended phase space and further enhanced the relation between charged AdS black holes and van der Waals liquid-gas systems. It has been shown that black holes are in general quite analogous to van der Waal fluids and exhibit the diverse behavior of different substances in everyday life. See the nice reviews \cite{Kubiznak3,Altamirano3,Dolan4,Kubiznak2} and references therein.

Here, we would like to focus on the van der Waals like behavior of charged AdS black holes in the $R+f(R)$ gravity with constant curvature~\cite{Moon98}, whose entropy, heat capacity and Helmholtz free energy was obtained in Ref.~\cite{Moon98}. Ref.~\cite{Chen} studied their $P-V$ criticality in the extended phase space and showed that van der Waals like behavior exists in the $P-V$ graph of $f(R)$ AdS black holes. Recently, we investigated their phase transition in the canonical ensemble and further showed that $T-S$ graphs of $f(R)$ AdS black holes exhibit reverse van der Waals like behavior \cite{xiong6}. In this paper, we would like to revisit the van der Waals like behavior from a totally different perspective. Namely, the two point correlation function. Studying the properties of $f(R)$ AdS black holes~\cite{Moon98}-\cite{Odintsov3} is of interest, because $f(R)$ gravity is one of modified gravity theories which successfully mimics the history of universe, especially the cosmic acceleration.

   On the other hand, investigating the properties of the two point correlation function is also intriguing itself. The two point correlation function is dual to the geodesic length according to the famous AdS/CFT correspondence \cite{ads1,ads2,ads3}. Interesting but intractable phenomena in strongly coupled system can be traced elegantly via the nonlocal observables such as two point correlation function, Wilson loop and entanglement entropy. Examples can be found in the researches of superconducting phase transition \cite{Superconductors1}-\cite{Ling}, the holographic thermalization \cite{Balasubramanian1}-\cite{Craps} and cosmological singularity \cite{Engelhardt,Engelhardt1}. Recently, the isocharges in the entanglement entropy-temperature plane was investigated in Ref. \cite{Johnson}, where both the critical temperature and critical exponent were proved to be exactly the same as the case of entropy-temperature plane. This finding is really intriguing and is attracting more and more attention \cite{Caceres}-\cite{Momeni}. Especially, it was proved in Ref. \cite{Nguyen} that the entanglement entropy-temperature plane obeys the equal area law just as $T-S$ curve does\cite{Spallucci}. In this paper, we would like to generalize these research to see whether the two point correlation function of $f(R)$ AdS black holes exhibits similar behavior. If it does, it would be a totally different perspective to observe the van der Waals like behavior of $f(R)$ AdS black holes other than $T-S$ graph and $P-V$ graph. To the best of our knowledge, this issue has not been covered in literature yet. As described above, the former literatures mainly focus on the entanglement entropy, including its van der Waals like behavior \cite{Johnson} and equal area law \cite{Nguyen}. Our paper mainly focus on the two point correlation function (not the entanglement entropy) of $f(R)$ AdS black holes. So the results presented in this paper is independent and will certainly contribute to the knowledge of both $f(R)$ gravity and AdS black holes.

The metrics of $f(R)$ AdS black holes and RN-AdS black holes look similar via rescaling of parameters. Then one may expect that the results presented in this paper can be compared with those of RN-AdS black holes by performing that rescaling. However, this is not the whole story. Similarities and differences coexist. So it is worth probing from two different perspectives. On the one hand, the more similarities they have, the more importance they gain. The amazing similarities imply that this black hole solution may serve as a bridge across the Einstein gravity and $f(R)$ gravity. The similarities will also call for further investigation which may shed light on some deeper physics which has not been disclosed yet. On the other hand, one will expect to search for the possible unique features that are different from the features in Einstein gravity. These differences are certainly of interest since the metrics look similar. In this paper, we will show that although the free energy changes with $b$, the first order phase transition temperature is the same for different $b$. If we interpret $q/\sqrt{b}$ as the rescaled charge that can be compared with that of RN-AdS black holes, then our result implies that the first order phase transition temperature does not vary with the rescaled charge $Q$ (at least it is true for the cases when $b$ varies). And it is quite different from the RN-AdS black holes whose first order phase transition temperature depends on $Q$.

    The organization of this paper is as follows. In Sec.~\ref{sec:2} we will have a brief review of critical phenomena of $f(R)$ AdS black hole. Two point correlation function of $f(R)$ AdS black holes will be investigated numerically in Sec.~\ref {sec:3}. Maxwell equal area law will be numerically checked in Sec.~\ref{sec:4}. Conclusions will be drawn in Sec.~\ref {sec:5}.

\section{A brief review of critical phenomena of charged AdS black holes in $f(R)$ gravity}
\label{sec:2}
  Ref.~\cite{Moon98} obtained in the $R+f(R)$ gravity with constant curvature scalar $R=R_0$ a charged AdS black hole solution, whose metric reads
\begin{equation}
ds^2=-N(r)dt^2+\frac{dr^2}{N(r)}+r^2(d\theta^2+\sin^2\theta d\phi^2),\label{1}
\end{equation}%
where
\begin{align}
N(r)&=1-\frac{2m}{r}+\frac{q^2}{br^2}-\frac{R_0}{12}r^2,\label{2}
\\
b&=1+f'(R_0).\label{3}
\end{align}%
Note that $b>0,R_0<0$.

This black hole solution is asymptotically AdS when one identify the curvature scalar as~\cite{Moon98}
\begin{equation}
R_0=-\frac{12}{l^2}=4\Lambda.\label{4}
\end{equation}%

The black hole ADM mass $M$ and the electric charge $Q$ are related to the parameters $m$ and $q$ respectively~\cite{Moon98}
\begin{equation}
M=mb,\;\;\; Q=\frac{q}{\sqrt{b}}.\label{5}
\end{equation}%

Ref.~\cite{Chen} reviewed its Hawking temperature, entropy and electric potential as
\begin{align}
T&=\frac{N'(r_+)}{4\pi}=\frac{1}{4\pi r_+}(1-\frac{q^2}{br_+^2}-\frac{R_0r_+^2}{4}).\label{6}
\\
S&=\pi r_+^2b.\label{7}
\\
\Phi&=\frac{\sqrt{b}q}{r_+}.\label{8}
\end{align}

\begin{figure*}
\centerline{\subfigure[]{\label{1a}
\includegraphics[width=8cm,height=6cm]{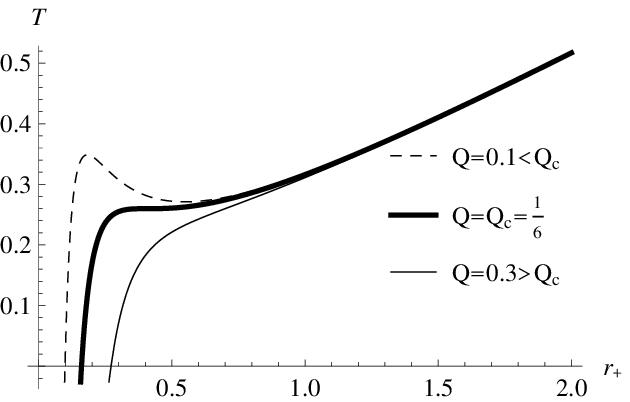}}
\subfigure[]{\label{1b}
\includegraphics[width=8cm,height=6cm]{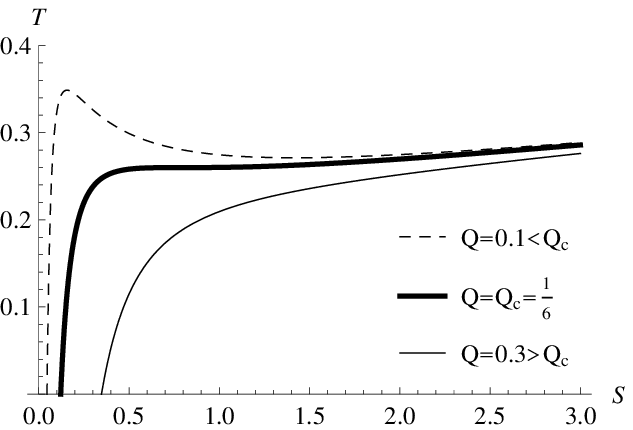}}}
 \caption{(a) $T$ vs. $r_+$ for $b=1.5, R_0=-12$ (b) $T$ vs. $S$ for $b=1.5, R_0=-12 $ [15]} \label{fg1}
\end{figure*}

In Ref. \cite{xiong6},  we investigated in detail the critical phenomena in the canonical ensemble. As shown in Fig.\ref{fg1}, both $T-S$ curve and $T-r_+$ curve show reverse van der Waals behavior (reverse denotes that at small $r_+$ or small $S$, $T\rightarrow0$ rather than $T\rightarrow\infty$) when $Q<Q_c$. The relevant critical quantities were derived as \cite{xiong6}
\begin{equation}
Q_c=\sqrt{\frac{-1}{3R_0}}, \;\;\;r_c=\sqrt{\frac{-2}{R_0}},\;\;\;S_c=\frac{-2b\pi}{R_0}.\label{9}
\end{equation}%
Numerical check of Maxwell equal area law for the cases $Q = 0.2Qc,0.4Qc,0.6Qc,0.8Qc$ was also carried out in $T-S$ graph. It was shown that the relative errors are amazingly small and the Maxwell equal area law holds for $T-S$ curve of $f(R)$ black holes \cite{xiong6}. The analytic expression of free energy was obtained as
\begin{equation}
F=\frac{R_0S^2+12\pi b(3b\pi Q^2+S)}{48\pi^{3/2}\sqrt{bS}}.\label{10}
\end{equation}%
It was shown that the classical swallow tails characteristic of first order phase transition appears in the case $Q<Q_c$ \cite{xiong6}.

\section{Two point correlation function of f(R) AdS black holes and its van der Waals like behavior}
\label{sec:3}
The equal time two point correlation function in the large $\Delta$ limit reads \cite{Balasubramanian61}
 \begin{equation}
\langle {\cal{O}} (t_0,x_i) {\cal{O}}(t_0, x_j)\rangle  \approx
e^{-\Delta {L}} ,\label{ll}
\end{equation}
where $L$ is the length of the bulk geodesic between the points $(t_0,
x_i)$ and $(t_0, x_j)$ on the AdS boundary while $\Delta$ denotes the conformal dimension of scalar operator $\cal{O}$ in the  dual field theory.

  Here, we choose two boundary points as $(\phi=\frac{\pi}{2},\theta=0)$ and $(\phi=\frac{\pi}{2},\theta=\theta_0)$ for simplicity. Parameterizing the trajectory with $\theta$, the proper length can be obtained as
\begin{eqnarray}
L=\int_0 ^{\theta_0}\mathcal{L}(r(\theta),\theta) d\theta,~~\mathcal{L}=\sqrt{\frac{\dot{r}^2}{N(r)}+r^2}, \label{12}
 \end{eqnarray}
where $\dot{r}=dr/ d\theta$.

Utilizing the Euler-Lagrange equation $\frac{\partial L}{\partial r}=\frac{d}{d\theta}\left(\frac{\partial L}{\partial r'}\right)$, one can derive the equation of motion for $r(\theta)$ as
\begin{equation}
2r(\theta)N(r)\ddot{r}(\theta)-[r(\theta)N'(r)+4N(r)]\dot{r}(\theta)^2-2N(r)^2r(\theta)^2=0.\label{13}
\end{equation}%
The boundary conditions can be fixed as
\begin{eqnarray}
r(0)= r_0, \dot{r}(0)=0.\label{14}
\end{eqnarray}
Solving the equation of motion (\ref{13}) constrained by the boundary condition (\ref{14}), one can obtain $r(\theta)$. However, the analytic expression is difficult to derive and we have to appeal for numerical methods.

It is worth mentioning that the geodesic length should be regularized by subtracting off the geodesic length in pure AdS with the same boundary region to avoid the divergence. We use $\delta L$ to denote the regularized  geodesic length, which can be calculated through the definition $\delta L\equiv L-L_0$. Note that $r_{AdS}(\theta)$ corresponding to entanglement entropy in pure AdS $L_0$ has been obtained analytically as $r_{AdS}(\theta)=l[(\frac{\cos \theta}{\cos \theta_0})^2-1]^{-1/2}$ \cite{Blanco, Casini}. Note that this formula can be applied to the case of $f(R)$ AdS black holes by introducing an effective cosmological constant and rescaling the AdS length (We appreciate the referee's deep insight for pointing out this issue for us). And the outcome is in accord with the numerical treatment.

\begin{figure*}
\centerline{\subfigure[]{\label{2a}
\includegraphics[width=8cm,height=6cm]{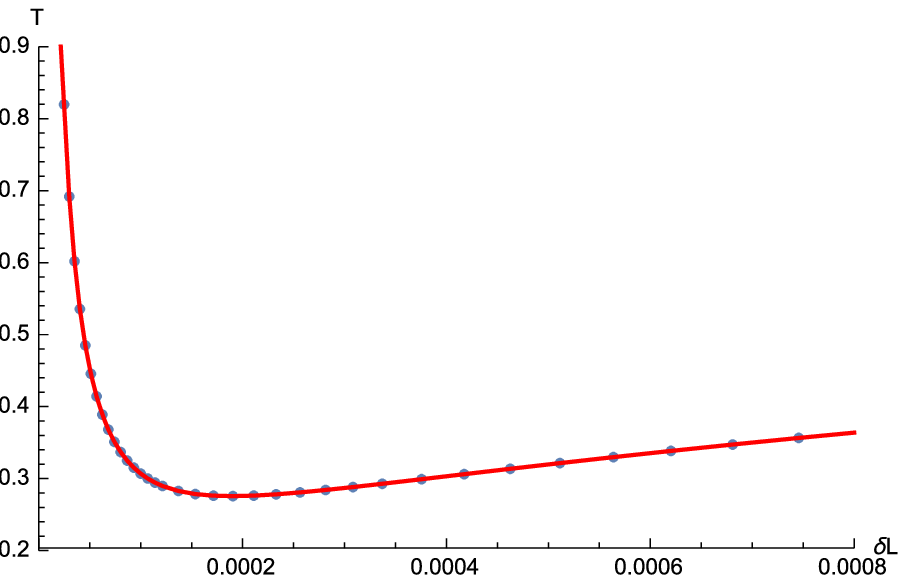}}
\subfigure[]{\label{2b}
\includegraphics[width=8cm,height=6cm]{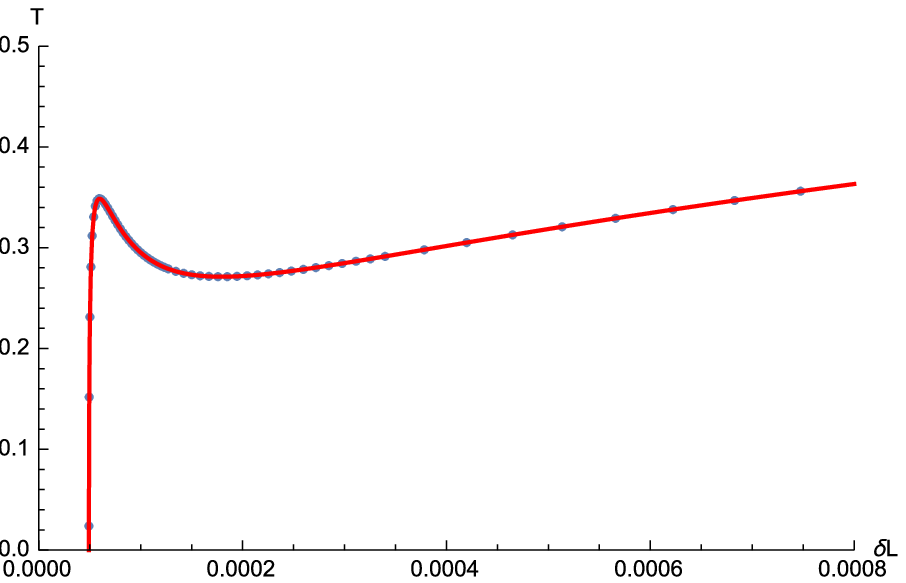}}}
\centerline{\subfigure[]{\label{2c}
\includegraphics[width=8cm,height=6cm]{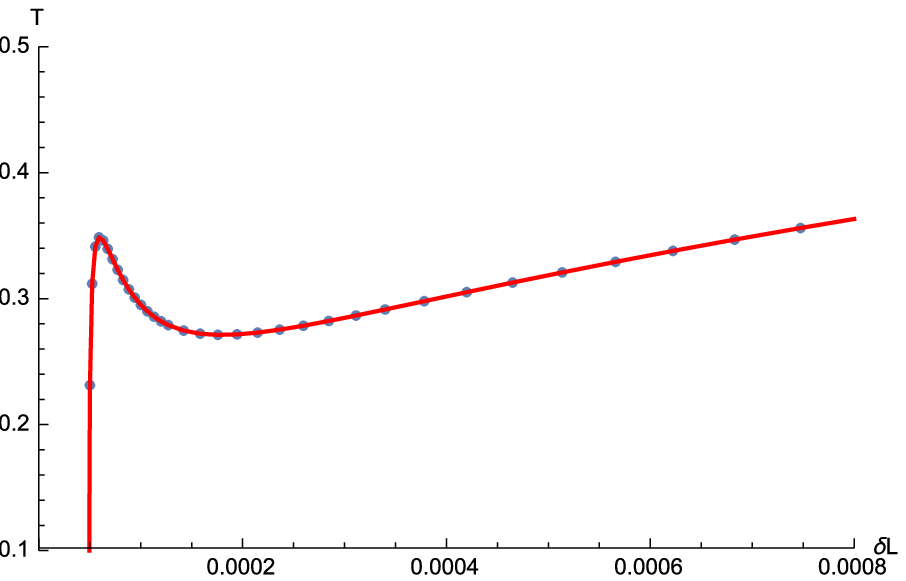}}
\subfigure[]{\label{2d}
\includegraphics[width=8cm,height=6cm]{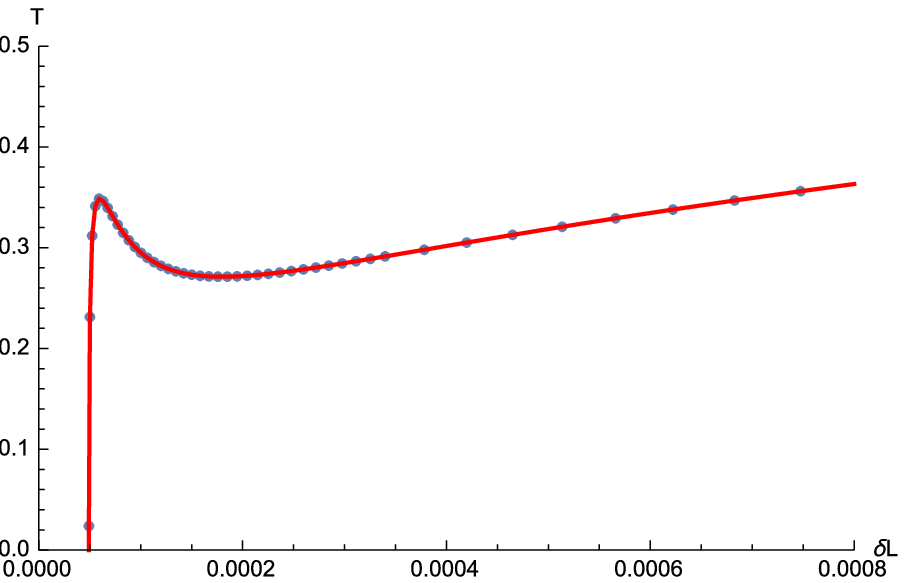}}}
 \caption{$T$ vs. $\delta L$ for $\theta_0=0.1$ (a) $b=1, Q=0$ (b) $b=0.5, Q=0.1$ (c) $b=1, Q=0.1$ (d) $b=1.5, Q=0.1$} \label{fg2}
\end{figure*}

\begin{figure*}
\centerline{\subfigure[]{\label{3a}
\includegraphics[width=8cm,height=6cm]{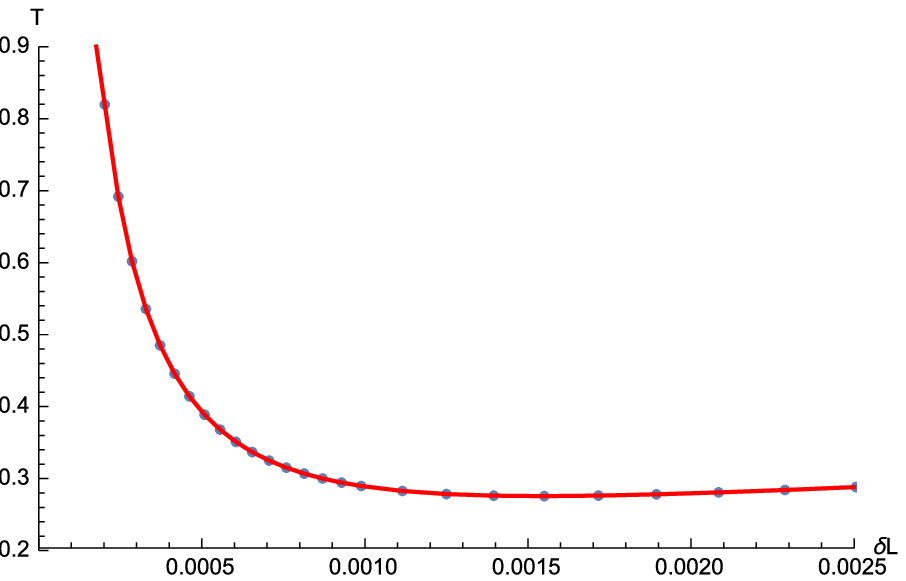}}
\subfigure[]{\label{3b}
\includegraphics[width=8cm,height=6cm]{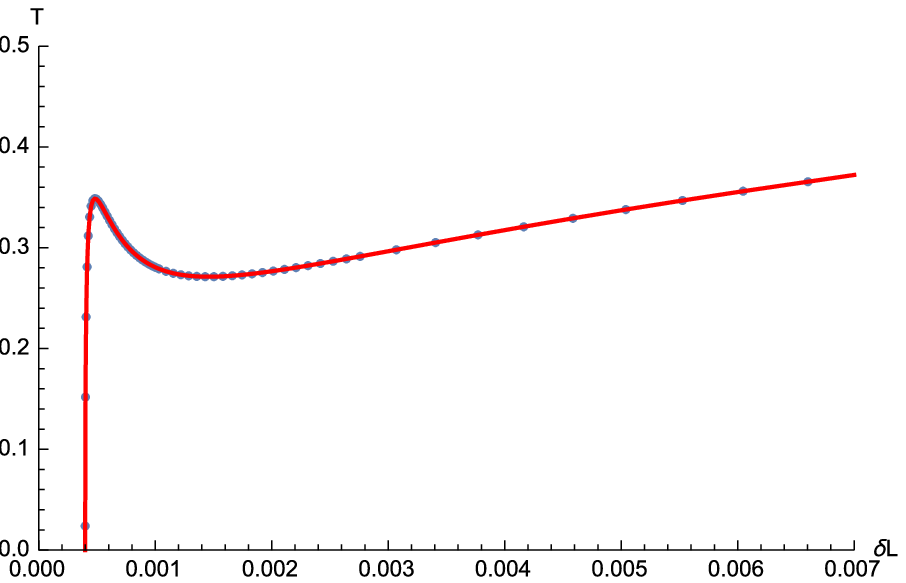}}}
\centerline{\subfigure[]{\label{3c}
\includegraphics[width=8cm,height=6cm]{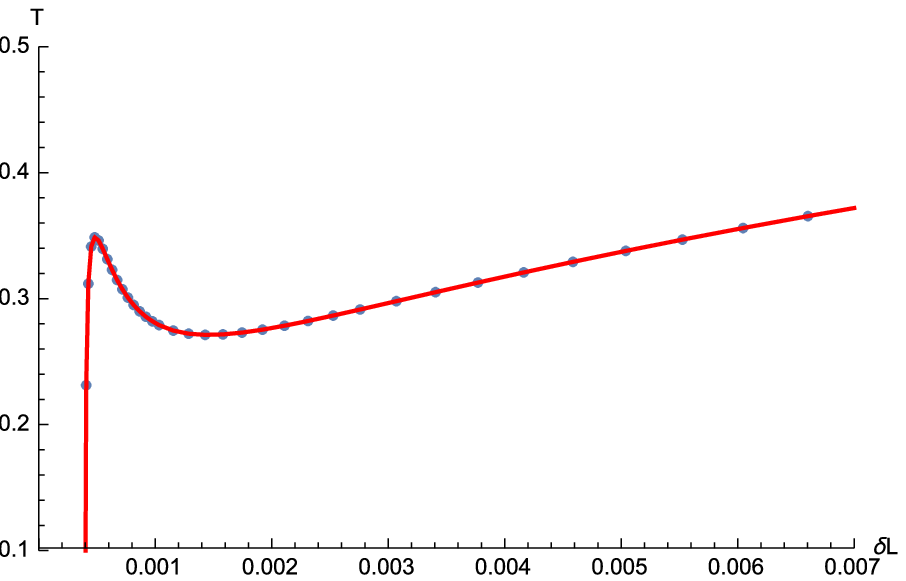}}
\subfigure[]{\label{3d}
\includegraphics[width=8cm,height=6cm]{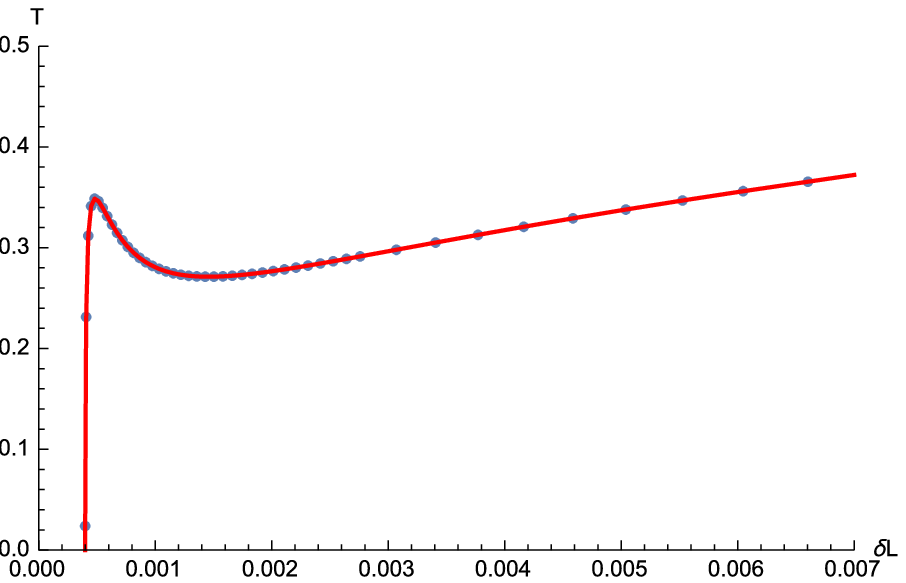}}}
 \caption{$T$ vs. $\delta L$ for $\theta_0=0.2$ (a) $b=1, Q=0$ (b) $b=0.5, Q=0.1$ (c) $b=1, Q=0.1$ (d) $b=1.5, Q=0.1$} \label{fg3}
\end{figure*}

To probe the effect of boundary region size on the phase structure, we choose $\theta_0=0.1, 0.2$ as two specific examples and we set the cutoff $\theta_c=0.099, 0.199$ respectively. On the other hand, we consider the case $b=0.5, 1, 1.5$ to investigate the effect of $f(R)$ gravity on the phase structure. For convenience, we set the AdS radius $l$ to be 1, which is equivalent to $R_0=-12$. According to Eq.(\ref{9}), $Q_c=1/6$. As we are interested in the possible reverse van der Waals behaviors, we would like to pay more attention to the case $Q<Q_c$. So we set the charge $Q$ to be $0.1$ in most of the cases in this paper. The case $Q=0$ is also probed for the purpose of comparison. As shown in Fig. \ref{fg2} and Fig. \ref{fg3}, the $T-\delta L$ graphs for $Q=0$ have only one inflection point while the graphs for $Q<Q_c$ have two inflection points. From the $T-\delta L$ graphs for $Q<Q_c$ , one can see clearly the reverse van der Waals like behavior. It is also shown that the effect of $b$ is so small that the $T-\delta L$ graphs seem the same at the first glance, although the specific numeric data points are different. On the other hand, the effect of boundary region size is quite obvious as reflected in the range of $\delta L$ axis.

\section{Numerical check of equal area law in $T-\delta L$ graph}
\label{sec:4}

To investigate the stability of black holes reflected in the $T-\delta L$ graph, one can introduce an analogous definition of specific heat as
\begin{equation}
C=T\frac{\partial \delta L}{\partial T}. \label{15}
 \end{equation}
Similar to the $T-S$ graph, one can find in the $T-\delta L$ graph that when $Q<Q_c$, both the large radius branch and the small radius branch are stable with positive specific heat while the medium radius branch is unstable with negative specific heat. Following a similar approach as that for $T-S$ graph \cite{Spallucci}, one can remove the unstable branch in $T-\delta L$ curve with a bar $T=T_*$ vertical to the temperature axis. Physically, $T_*$ denotes the Hawking temperature corresponding to the first order phase transition. The analogous Maxwell equal area law in the $T-\delta L$ graph may be written as
\begin{equation}
T_*\times(\delta L_3-\delta L_1)=\int^{\delta L_3}_{\delta L_1}Td\delta L,\label{16}
\end{equation}%
where $\delta L_1$, $\delta L_2$, $\delta L_3$ denote the three values of $\delta L$ at $T=T_*$ with the assumption that $\delta L_1<\delta L_2<\delta L_3$.

\begin{figure*}
\centerline{\subfigure[]{\label{4a}
\includegraphics[width=8cm,height=6cm]{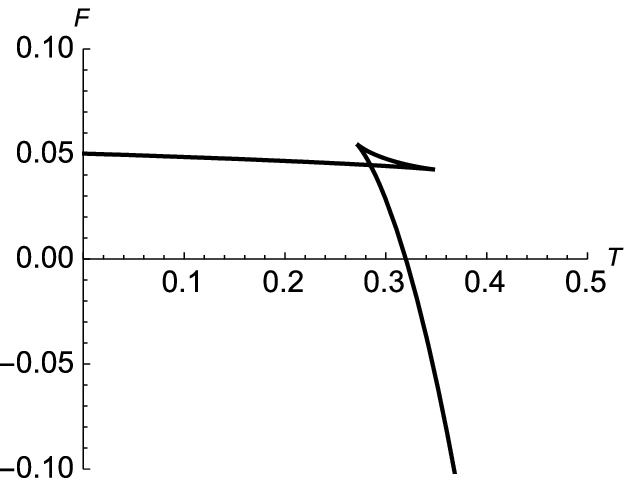}}
\subfigure[]{\label{4b}
\includegraphics[width=8cm,height=6cm]{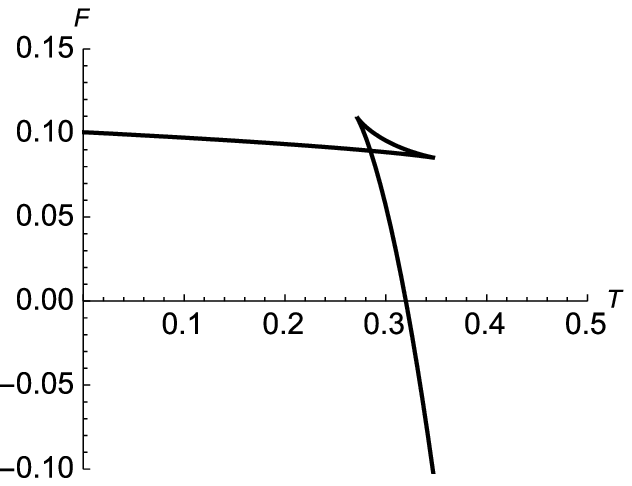}}}
\centerline{\subfigure[]{\label{4c}
\includegraphics[width=8cm,height=6cm]{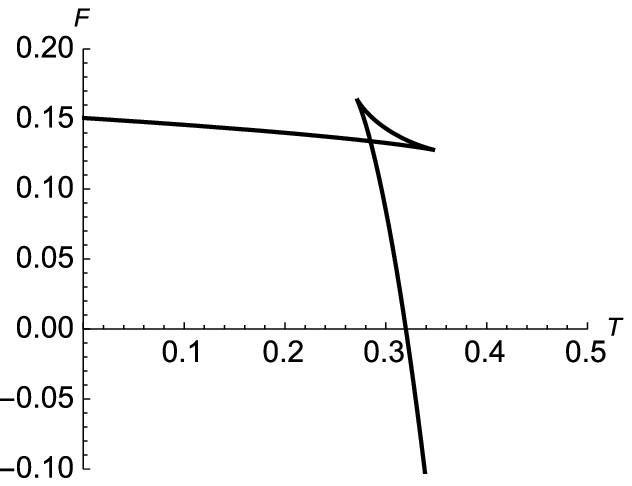}}}
 \caption{$F$ vs. $T$ for $Q=0.1,R_0=-12$ (a) $b=0.5$ (b) $b=1$ (c) $b=1.5$} \label{fg4}
\end{figure*}

$T_*$ can be determined utilizing the free energy analysis. One can find $T_*$ from the intersection point of two branches in the $F-T$ graph. The free energy graphs for different choice of $b$ is plotted in Fig. \ref{fg4}. From Fig. \ref{fg4}, one can see clearly that the values of free energy are influenced by the parameter $b$ showing the impact of $f(R)$ gravity. To one's surprise, the effect of $b$ on $T_*$ is very little at the first glance at Fig. \ref{fg4}. We further obtain numerical result of $T_*$ to be $0.2847050173$ for the three choices of $b$, namely, 0.5, 1, 1.5, implying that the first order phase transition temperature $T_*$ is the same at least to the order of $10^{-10}$ for different choices of $b$. Considering the fact that the $f(R)$ AdS black holes reduce to RN-AdS black holes when $b=1$, the result we obtain here is very interesting for it implies that $f(R)$ gravity does not influence the first order phase transition temperature. Our result further supports the former finding that charged $f(R)$ AdS black holes behave much like RN-AdS black holes. In former literature, it was reported in the reduced parameter space that charged $f(R)$ AdS black holes share the same equation of state~\cite{Chen}, the same coexistence curve~\cite{xiong5} and the same molecule number density difference~\cite{xiong5} with RN-AdS black holes. This can be attributed to the observation that these two black hole metrics are identical when the charge is rescaled and an effective cosmological constant is defined for $f(R)$ AdS black holes (We appreciate the referee's deep insight for pointing out this issue for us). It is interesting that these two black holes in two different gravity theories ($f(R)$ gravity and Einstein gravity) share so much similarities in their metrics and other behaviors.

With $T_*$ at hand, we calculate numerically the values of the left-hand side and right-hand side of Eq.(\ref{16}) for the cases $\theta_0=0.1$ and $\theta_0=0.2$ respectively to check whether the analogous Maxwell equal area law holds for $T-\delta L$ curve. As can be witnessed from Table. \ref{tb1}, the relative errors for both cases are so small that we can safely draw the conclusion that the analogous Maxwell equal area law holds for $T-\delta L$ curve of $f(R)$ AdS black holes.

\begin{table}[!h]
\tabcolsep 0pt
\caption{Numerical check of Maxwell equal area law in $T-\delta L$ graph for $R_0=-12, Q=0.1, b=1.5$}
\vspace*{-12pt}
\begin{center}
\def\temptablewidth{1\textwidth}
{\rule{\temptablewidth}{2pt}}
\begin{tabular*}{\temptablewidth}{@{\extracolsep{\fill}}c||c||c||c||c||c||c}
$\theta_0$ & $\delta L_1$ &$\delta L_2$ &$\delta L_3$ &$T_*(\delta L_3-\delta L_1)$ & $\int^{\delta L_3}_{\delta L_1}Td\delta L$ & relative error \\   \hline
 0.1      & 0.0000516933       &       0.0001149030 & 0.0002999516 & 0.00007068 & 0.00007099 & $0.4367 \%$ \\
 0.2      & 0.0004189403       &       0.0009334939 & 0.0024361593 & 0.00057431 & 0.00057684 & $0.4386\%$

       \end{tabular*}
       {\rule{\temptablewidth}{2pt}}
       \end{center}
       \label{tb1}
       \end{table}

\begin{figure*}
\centerline{\subfigure[]{\label{5a}
\includegraphics[width=8cm,height=6cm]{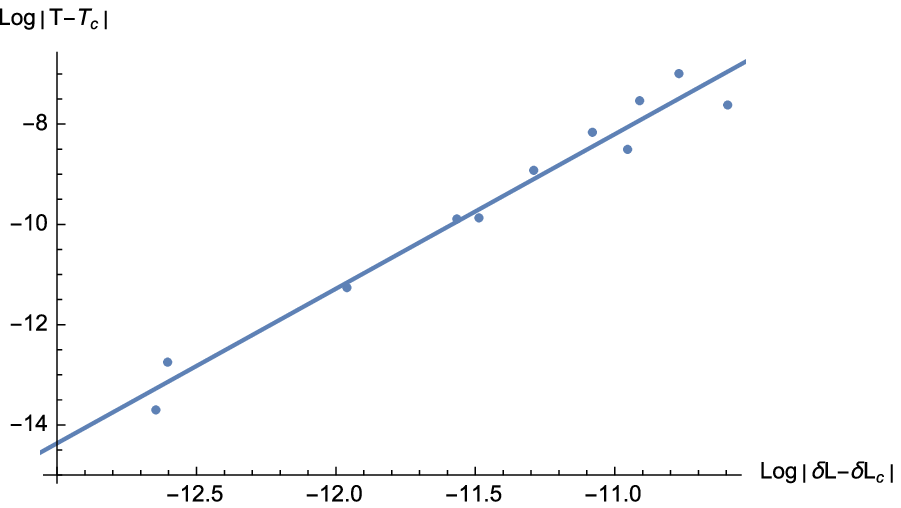}}
\subfigure[]{\label{5b}
\includegraphics[width=8cm,height=6cm]{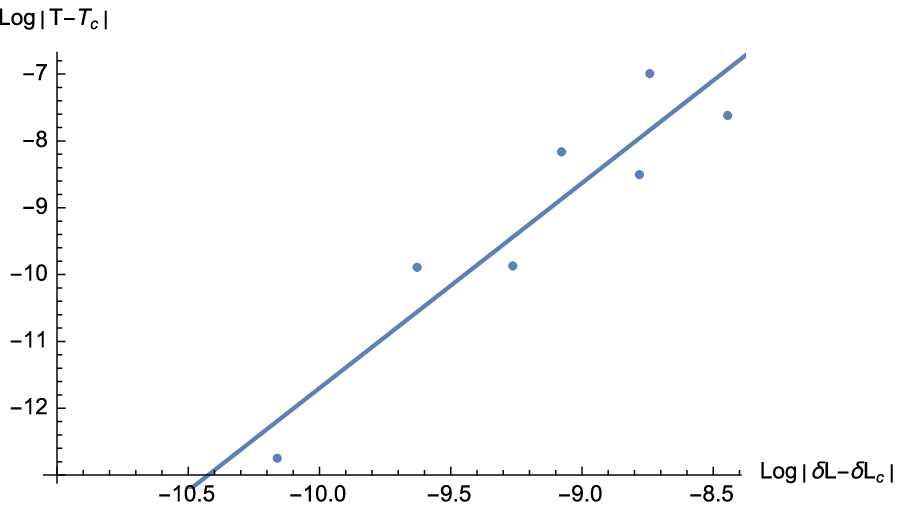}}}
 \caption{$\log\mid T-T_c\mid$ vs. $ \log\mid\delta L-\delta L_c\mid$ for $R_0=-12,b=1.5,Q=Q_c=1/6$ (a) $\theta_0=0.1$ (b) $\theta_0=0.2$} \label{fg5}
\end{figure*}

Utilizing Eqs.(\ref{6}) and (\ref{9}), one can easily derive the critical temperature as
\begin{equation}
T_c=\frac{\sqrt{-R_0}}{3\sqrt{2}\pi}. \label{17}
 \end{equation}
Substituting $R_0=-12$ into the above equation, one can obtain $T_c=0.2598989337$. Note that it is independent of $\theta_0$. Then one can adopt the interpolating functions obtained from the numeric result to obtain $\delta L_c$. For the case $\theta_0=0.1$, $\delta L_c=0.0001337812$ while for the case $\theta_0=0.2$, $\delta L_c=0.0010728619$. The relation between $ \log\mid T -T_c\mid$ and $\log\mid\delta L-\delta L_c\mid $ is plotted for the cases $\theta_0=0.1$ and $\theta_0=0.2$ in Fig. \ref{5a} and \ref{5b} respectively. Note that the points are chosen neighboring the critical point from the $T-\delta L$ graph corresponding to $Q=Q_c=1/6$. And the fitting function for these two cases are obtained as
\begin{equation}
\log\mid T-T_c\mid=\begin{cases}
25.7008 + 3.08202     \log\mid\delta L-\delta L_c\mid,&$for$~\theta_0=0.1,\\
18.9784 + 3.06772   \log\mid\delta L-\delta L_c\mid,&  $for$ ~\theta_0=0.2.\\
\end{cases}
\end{equation}
 From above, one can see clearly that the slope is around 3, implying that the critical exponent (defined through $C\sim \mid T-T_c\mid ^{-\alpha}$) is about $2/3$. Our result of critical exponent for the analogous specific heat related to the $\delta L$ is in accordance with those in former literatures of specific heat related to the thermal entropy \cite{Chamblin2} or entanglement entropy \cite{Johnson}.

\section{Conclusions}
\label{sec:5}
 In this paper, we focus on the two point correlation function of $f(R)$ AdS black hole. First, we choose two boundary points as $(\phi=\frac{\pi}{2},\theta=0)$ and $(\phi=\frac{\pi}{2},\theta=\theta_0)$ and obtain the proper length by parameterizing the trajectory with $\theta$. Then utilizing the Euler-Lagrange equation, we derive the equation of motion for $r(\theta)$. Solving the equation of motion constrained by the boundary condition, we obtain $r(\theta)$ numerically. Moreover, we regularize the geodesic length by subtracting off the geodesic length in pure AdS with the same boundary region.

Second, to probe the effect of boundary region size on the phase structure, we choose $\theta_0=0.1, 0.2$ as two specific examples and we set the cutoff $\theta_c=0.099, 0.199$ respectively. On the other hand, we consider the case $b=0.5, 1, 1.5$ to investigate the effect of $f(R)$ gravity on the phase structure. The case $Q<Q_c$ are plotted for different cases. From the $T-\delta L$ graphs, one can see clearly the reverse van der Waals like behavior. It is also shown that the effect of $b$ is so small that the $T-\delta L$ graphs seem the same at the first glance, although the specific numeric data points are different. On the other hand, the effect of boundary region size is quite obvious as reflected in the range of $\delta L$ axis.

Third, we introduce an analogous definition of specific heat to investigate the stability of black holes reflected in the $T-\delta L$ graph. Similar to the $T-S$ graph, one can find in the $T-\delta L$ graph that when $Q<Q_c$, both the large radius branch and the small radius branch are stable with positive specific heat while the medium radius branch is unstable with negative specific heat. We remove the unstable branch in $T-\delta L$ curve with a bar $T=T_*$ vertical to the temperature axis, where $T_*$ denotes the Hawking temperature corresponding to the first order phase transition. We carry out free energy analysis to determine $T_*$. It is shown that the values of free energy are influenced by the parameter $b$ showing the impact of $f(R)$ gravity. However, to one's surprise, the numerical result of $T_*$ to be $0.2847050173$ for all the three choices of $b$, namely, 0.5, 1, 1.5, implying that the first order phase transition temperature $T_*$ is the same at least to the order of $10^{-10}$ for different choices of $b$. Considering the fact that the $f(R)$ AdS black holes reduce to RN-AdS black holes when $b=1$, the result we obtain here is very interesting for it implies that $f(R)$ gravity does not influence the first order phase transition temperature. Our result further supports the finding that charged $f(R)$ AdS black holes behave much like RN-AdS black holes reported in former literature, where charged $f(R)$ AdS black holes were shown to share the same equation of state~\cite{Chen}, the same coexistence curve~\cite{xiong5} and the same molecule number density difference~\cite{xiong5} with RN-AdS black holes in the reduced parameter space. We also check the analogous equal area law numerically. It is shown that the relative errors for both cases are so small that we can safely draw the conclusion that the analogous Maxwell equal area law holds for $T-\delta L$ curve of $f(R)$ AdS black holes.

Last but not the least, we plot the relation between $ \log\mid T -T_c\mid$ and $\log\mid\delta L-\delta L_c\mid $ for the cases $\theta_0=0.1$ and $\theta_0=0.2$ and obtain the fitting functions for these two cases. It is shown that the slope is around 3, implying that the critical exponent (defined through $C\sim \mid T-T_c\mid ^{-\alpha}$) is about $2/3$. Our result of critical exponent for the analogous specific heat related to the $\delta L$ is in accordance with those in former literatures of specific heat related to the thermal entropy \cite{Chamblin2} or entanglement entropy \cite{Johnson}.

To summary, the $T-\delta L$ graph of $f(R)$ AdS black holes exhibits the reverse van der Waals like behavior just as the $T-S$ graphs do. And the analogous Maxwell equal area law holds for $T-\delta L$ graph. Moreover, the critical exponent for the analogous specific heat related to the $\delta L$ is shown to be the same as those of specific heat related to the thermal entropy or entanglement entropy. So the two point correlation function may serves as an alternative perspective to observe van der Waals like behavior of f(R) AdS black holes.

 \section*{Acknowledgements}
 The authors want to express their sincere gratitude to both the editor and the referee for their joint effort to improve the quality of this paper significantly. This research is supported by National Natural Science Foundation of China (Grant No. 11605082), and in part supported by Natural Science Foundation of Guangdong Province, China (Grant Nos. 2016A030310363, 2016A030307051, 2015A030313789). Xiao-Xiong Zeng is supported by the National Natural Science Foundation of China (Grant No. 11405016) and  Natural Science Foundation of  Education Committee of Chongqing (Grant No. KJ1500530).

\end{document}